\documentclass[aps,prl,twocolumn, superscriptaddress,showpacs]{revtex4}
\usepackage{graphicx}
\usepackage{psfrag}
\usepackage{subfigure}
\usepackage{verbatim}
\usepackage{color}
\usepackage{amsmath,amssymb}
\newcommand{\bea}{\begin{eqnarray}}
\newcommand{\eea}{\end{eqnarray}}

\begin{document}


\title{From damage percolation to crack nucleation through finite size criticality}



\author{Ashivni Shekhawat}
\affiliation{LASSP, Physics Department, Clark Hall, Cornell University, Ithaca, NY 14853-2501}
\author{Stefano Zapperi}
\affiliation{Consiglio Nazionale delle Ricerche-IENI, Via R. Cozzi 53, 20125 Milano, Italy}
\affiliation{ISI Foundation, Via Alassio 11C, 10126 Torino, Italy}
\author{James P.~Sethna}
\affiliation{LASSP, Physics Department, Clark Hall, Cornell University, Ithaca, NY 14853-2501}


\date{\today}

\begin{abstract}
We present a unified theory of fracture in disordered brittle media that
reconciles apparently conflicting results reported
in the literature. Our renormalization group based approach yields a
phase diagram in which the percolation fixed point, expected for
infinite disorder, is unstable for finite disorder and flows to a
zero-disorder nucleation-type fixed point, thus showing that fracture
has mixed first order and continuous character. In a region of intermediate disorder and
finite system sizes, we predict a crossover with mean-field avalanche scaling.
We discuss intriguing connections to other phenomena where
critical scaling is only observed in finite size systems and disappears in the
thermodynamic limit.
\end{abstract}





\maketitle
Brittle fracture in disordered media intertwines two phenomena that seldom
coexist, namely, nucleation and critical fluctuations. 
The usual dichotomy of thought between nucleated and continuous transitions
makes the study of fracture interesting. Even more intriguing is the fact 
that crack nucleation happens at zero stress in the thermodynamic limit: 
smaller is stronger and larger is weaker. This makes the existence 
of critical fluctuation in the form of clusters and avalanches of all
sizes even more mysterious. 
What kind of critical point
governs a phase transition that happens at zero applied field (stress) in the 
thermodynamic limit, and what is the universality class of such a transition?  
How do self-similar
clusters, extremely rough crack surfaces, and scale invariant avalanches ultimately give rise 
to sharp cracks and localized growth? 
These questions have been addressed previously via a host of different theories, 
such as those based on percolation and multifractals~\cite{roux1988,hansen2003,hansen1991,hansen2006}, 
spinodal modes and mean-field criticality~\cite{zapperi1997}, and 
classical nucleation~\cite{duxbury1987,duxbury1986,shekhawat2012,chakrabarti}.  
In this Letter, we present a theoretical framework based on 
the renormalization group and crossover scaling that unifies 
the seemingly disparate descriptions of fracture into one consistent framework.
\par

\par
Fracture in disordered media is the result of a complex interplay between quenched heterogeneities
and long-range stress fields leading  to diffuse damage throughout the sample,
and local stress concentration favoring the formation of sharp localized cracks. The self-affine morphology of cracks
\cite{bonamy11}, the power-law statistics of avalanche precursors \cite{hemmer1992,petri94,garcimartin97,salminen02}, 
and the scale
dependence of the failure strength distribution \cite{alava2006,bazant04b,harlow1978}, all result from 
this competition.
Disordered fracture can be understood in the limit of infinitesimal as 
well as infinite disorder. Infinitesimal disorder means perfect crystalline material with just a few isolated defects (say a missing 
atom or a micro-crack). 
In this limit, fracture statistics can be understood as a nucleation 
type first order phase transition~\cite{duxbury1986,duxbury1987,chakrabarti,shekhawat2012}.
In the limit of infinite disorder, stress concentration
becomes irrelevant and fracture progresses via uncorrelated percolation-like damage~\cite{roux1988,hansen2003}. 
This mapping to percolation theory becomes rigorously valid  when the disorder distribution is not normalizable (or
\emph{very broad}, in the language of multifractals)~\cite{roux1988}.
The situation is more interesting at intermediate disorder, where unlike typical first order transitions,
crack nucleation is preceded 
by avalanches with power-law distributions  and mean-field exponents  \cite{hansen94,zapperi1997,zapperi05,zapperi05b}, sometimes
interpreted as a signature of a spinodal point~\cite{zapperi1997}.
Our renormalization  group based theory unifies the above
descriptions into a single phase diagram.
\par
\begin{figure}[tbp]
\begin{center}
\subfigure[A fuse network]{\includegraphics[width=0.24\textwidth,angle=0]{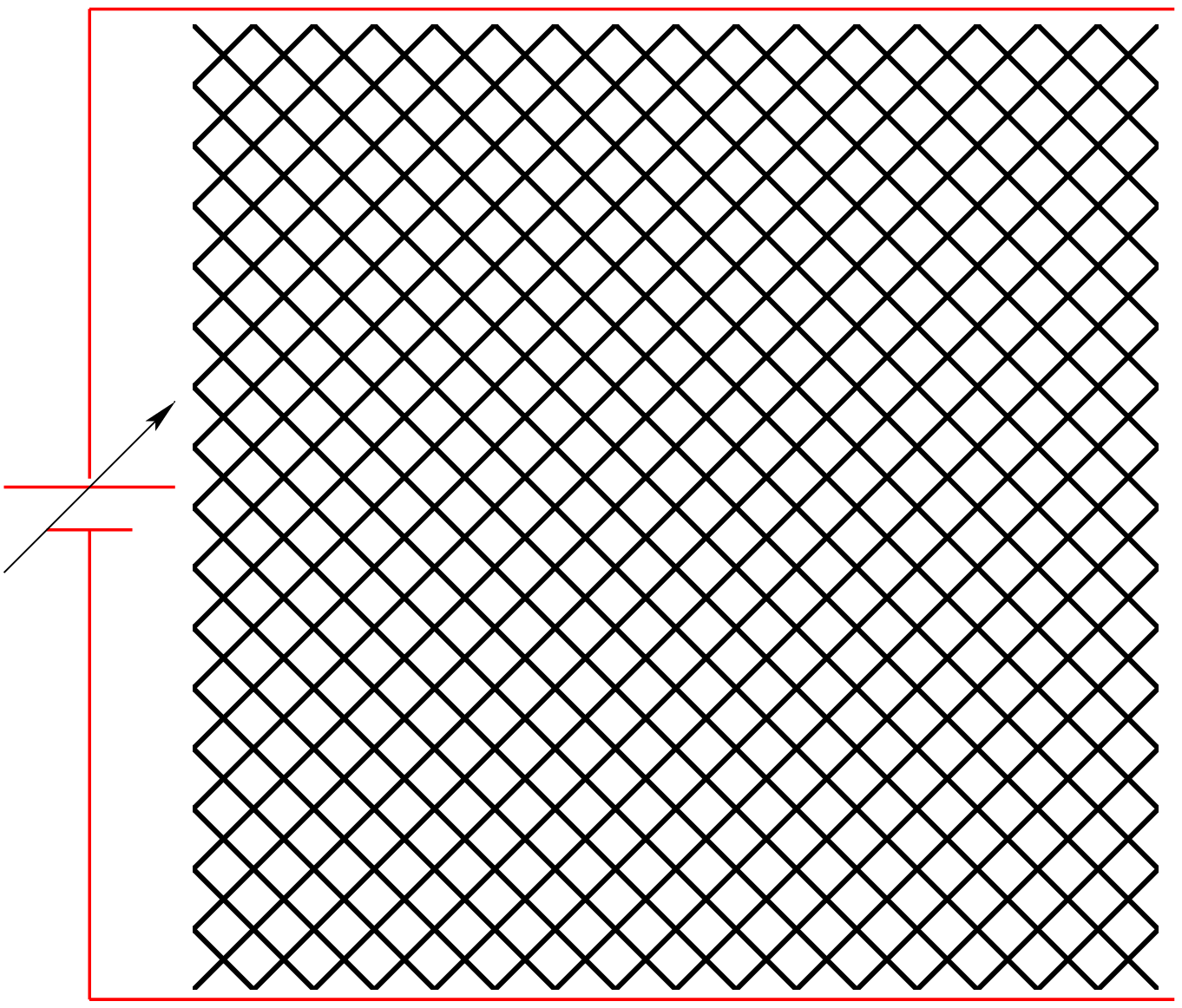}}
\subfigure[$\beta = 0.03$]{\includegraphics[width=0.21\textwidth,angle=0]{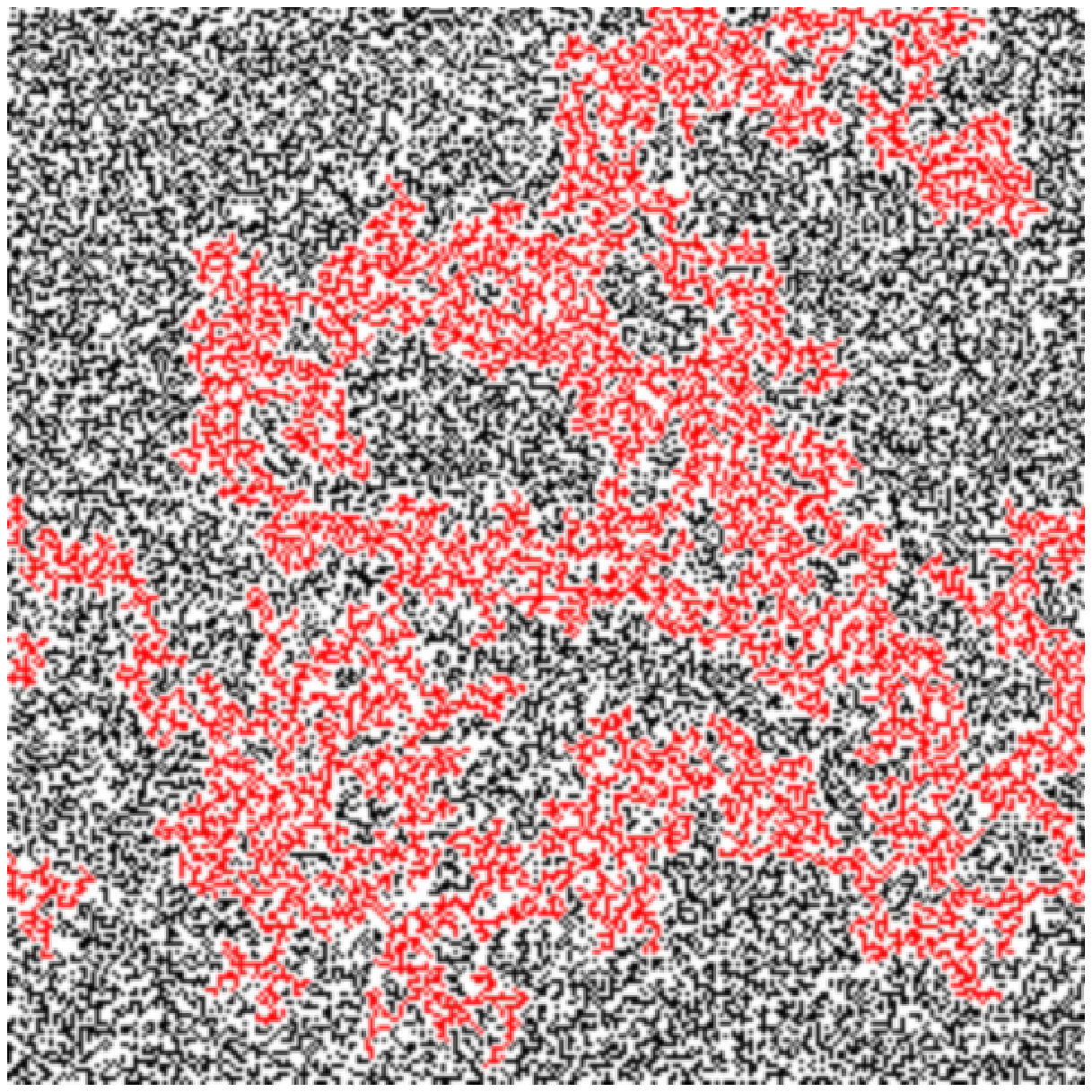}}
\end{center}
\begin{center}
\subfigure[$\beta = 0.5$ ]{\includegraphics[width=0.21\textwidth,angle=0]{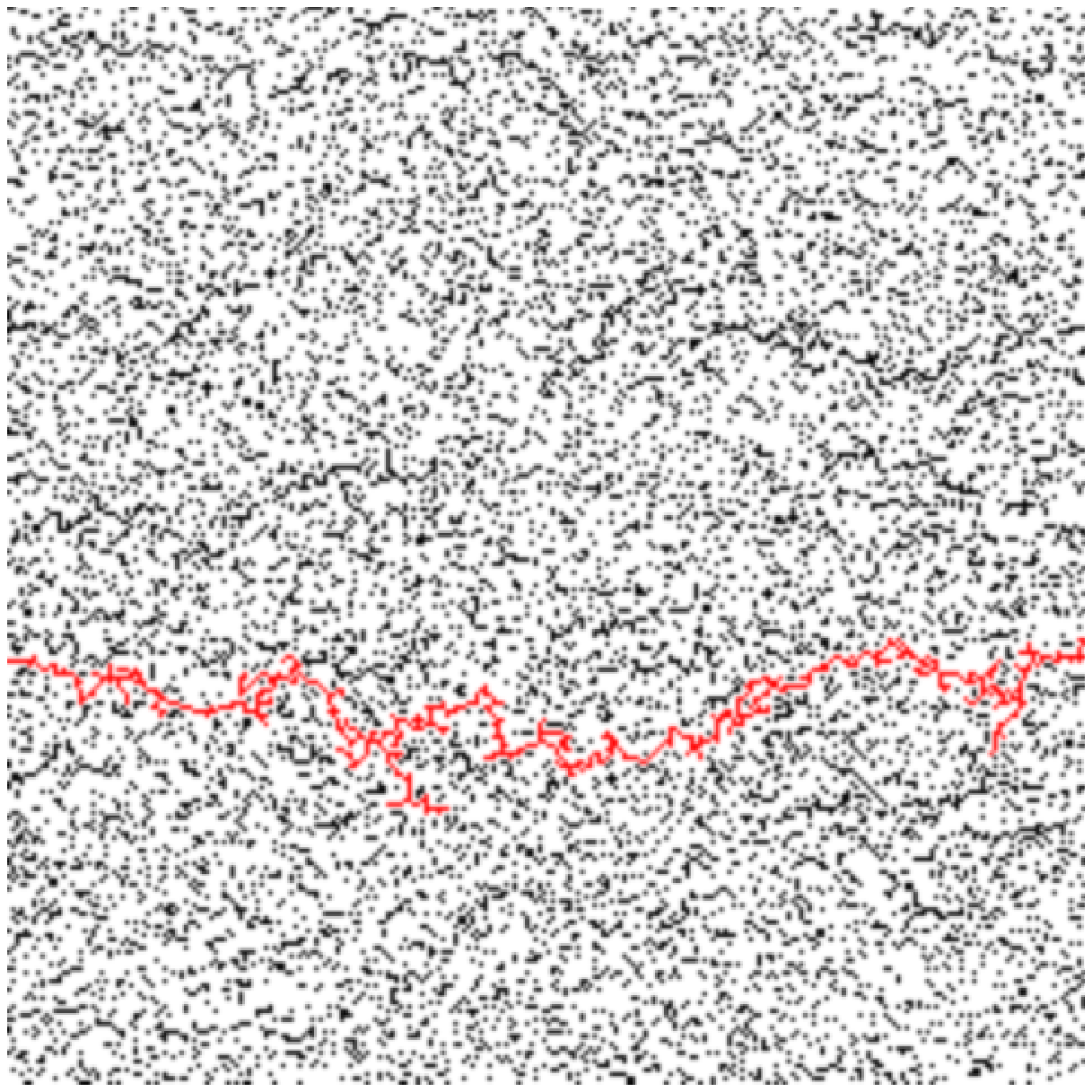}}
\subfigure[$\beta = 3.0$ ]{\includegraphics[width=0.21\textwidth,angle=0]{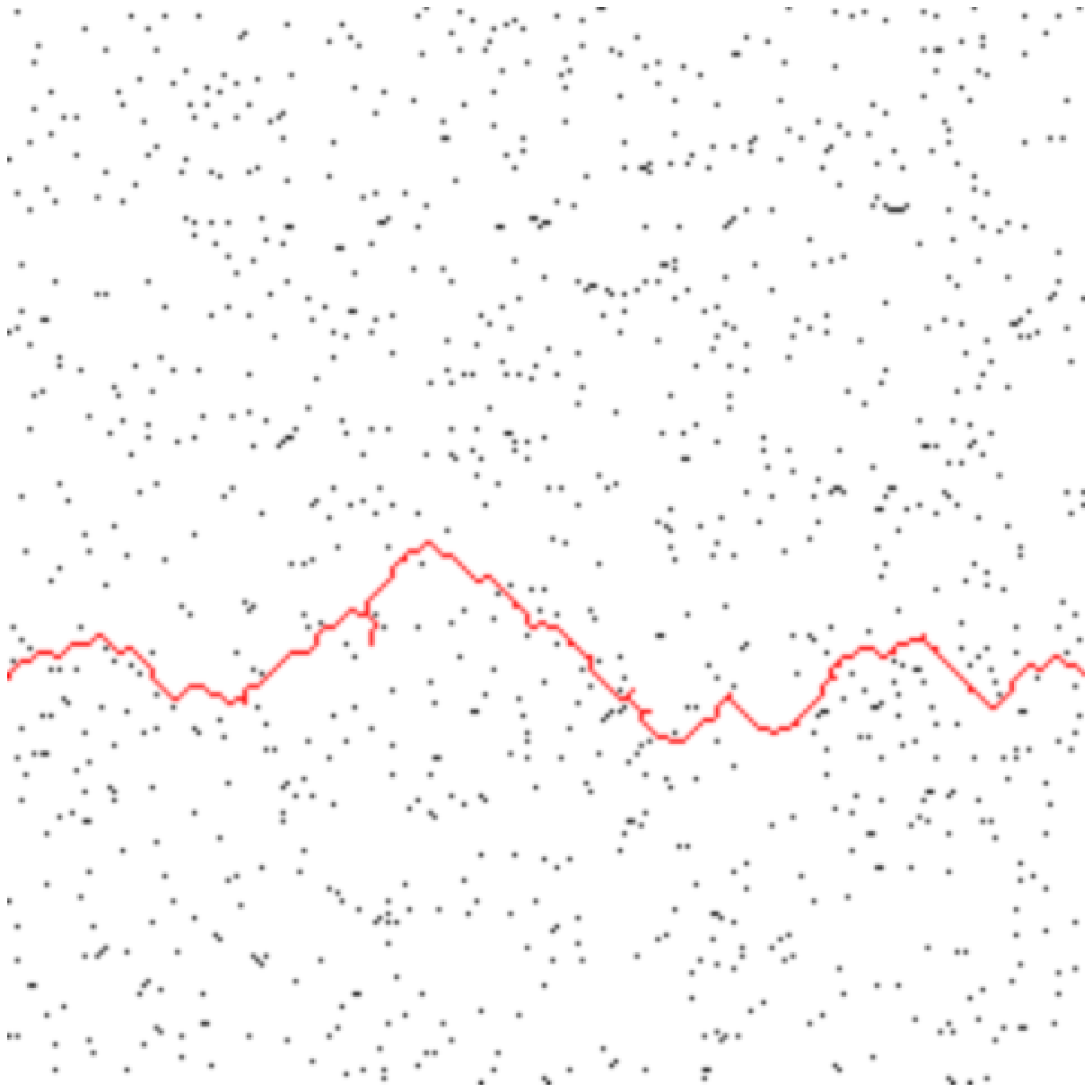}}
\end{center}
\caption{{\bf Fuse network model. a).} Schematic of a fuse network. Periodic boundary conditions
are used in the horizontal direction. {\bf b-d).} Fractured sample for various values 
of the parameter $\beta$; the spanning cluster (or crack) is colored red. 
There is a smooth crossover from percolation-like behavior for 
small $\beta$ to nucleated cracks at large $\beta$.
} 
\label{fig:Network}
\end{figure}

We use a 2D fuse network to model disordered brittle materials. 
A description of the disordered fuse network model that we study can be found in any 
number of references~\cite{zapperi1997,nukala2003,hansen2003,alava2006,kahng88,hansen1991,hansen2006}. 
Briefly, we consider a periodic network of fuses arranged in a square lattice of size $L$ tilted by 45$^\circ$ 
(the so-called `diamond lattice', figure 1a). Each fuse is assigned a quenched current threshold from a common 
distribution with a cumulative distribution function $F(\cdot)$. 
If the current through a fuse exceeds its threshold, then the fuse is burned and 
is removed from the network i.e.,~its conductance is set to zero. 
The current through the network is ramped quasi-statically, and fuses are burned one at a time
until the network becomes non-conducting, at which point the network is said to be fractured. 
We assign thresholds between 0 and 1, specifically we take $F(x) = x^\beta,\ \beta > 0$.
This form of distribution of 
thresholds serves as model for a generic distribution with a power-law tail at the origin, and has been 
studied widely~\cite{zapperi1997,hansen2003,hansen1991,hansen2006}. In this model the limit $\beta \to 0$
corresponds to infinite disorder, while the limit $\beta \to \infty$ corresponds to 
infinitesimal disorder.
Figure~\ref{fig:Network} shows a schematic of an undamaged fuse network (\ref{fig:Network}a),
and realizations of fractured networks for various values of the parameter $\beta$. Notice 
how the damage looks percolation-like for small $\beta$ while a single crack appears for 
large $\beta$.

\par
We begin by arguing that crack-tip stress concentration is a relevant perturbation to the infinite disorder 
percolation critical point. Our assertion implies that percolation-like behavior is a finite-size crossover effect.
This is consistent with recent results showing that even an arbitrarily small cutoff in the 
threshold distribution (at $\beta = 0$) leads to a crossover away from percolation at large system sizes~\cite{Hansen2012}.
We calculate self-consistent upper and lower bounds for the stress and damage fraction at failure,
and show that all of these quantities vanish in the limit of large $L$. This will establish that percolation 
cannot be the dominant behavior for large $L$, since percolation demands that the damage fraction be finite.
Let $\sigma_f,\ \phi_f$ be the stress and the damage fraction
at failure, respectively. The lower bound on both quantities is trivially equal to 0. The upper bound is obtained self-consistently.
Let us assume that $\phi_f < \phi_f^+ \ll 1$, where $\phi_f^+$ is an upper bound on $\phi_f$, and 
similarly $\sigma_f < \sigma_f^+ \ll 1$.
Let, if possible, the damage be percolation-like, so that $\phi_f^+ = F(\sigma_f^+)$
\footnote{A calculation based on effective medium theory yields $\phi = F(\sigma/(1 - \phi^2)) \sim 
F(\sigma) + \mathcal{O}(\sigma\phi^2)$. We ignore the higher order terms.}.
The stress at the tip of a crack of length $l$ (lattice units) is given by $\sigma_{tip}(l) \approx \sigma_f^+(1 + \alpha \sqrt{l})$, 
where $\alpha$ is a lattice dependent constant. Thus, the length of a \emph{critical crack} at a given stress
and damage fraction is $l_{cr}(\sigma_f^+) \sim 1/(\sigma_f^+)^2\alpha^2 +\ h.o.t$. The probability 
that a critical crack forms at a given lattice site is at least $F(\sigma_f^+)^{l_{cr}}$. Since there are $L^2$ sites in the 
lattice, the probability of 1 such crack appearing on the entire lattice is at least $L^2 F(\sigma_f^+)^{l_{cr}}$~\cite{duxbury1987}.
At the failure stress this probability is 1, thus $\sigma_f^+$ can be obtained by solving $L^2 F(\sigma_f^+)^{l_{cr}(\sigma_f^+)} = 1$.
It can be proved that the solution $\sigma_f^+(L) \to 0$ as $L \to \infty$, thus, $\phi_f^+ = F(\sigma_f^+) \to 0$.
\par
We have established that percolation is unstable to nucleation, however, the crossover length is expected to be 
rather large. The reason for this effect is that $\phi_f^+(L)$ decays very slowly with $L$. 
This slow decay is expected to also manifest itself in the form of large finite size effects.
The rate of 
decay obviously depends on $F(\cdot)$, for $F(x) = x^\beta$ 
one can show that $\phi_f^+(L) \sim (\beta/2\log L)^{\beta/2}$. 
More sophisticated estimates that account for stress concentration during the growth of the critical crack,
as opposed to percolation-like growth assumed here, yield similar results. The convergence becomes extremely slow as $\beta$ approaches
0, meaning that percolation threshold will be reached before nucleation 
of the critical crack for increasing larger system sizes. This is consistent with the previous studies that 
found that the fuse network can be mapped onto a percolation 
problem in the limit of $\beta \to 0$~\cite{hansen2003}. 
However, one should note the subtle point that order of limits matters since percolation
is ultimately unstable to nucleation at any $\beta$.
\par
The avalanche behavior associated with fracture can be understood via a simple argument. The argument 
is valid in the vicinity of the critical point and breaks down for very large $L$. Consider an avalanche that starts
with a bond breaking at a stress $\sigma\ (\ll 1)$ and damage fraction $\phi\ (\ll 1)$. Linear elasticity predicts that 
the change in the stress field due to the breaking of the bond, $c(r,\sigma)$, decays as $c(r,\sigma) \sim \sigma/r^2$ (ignoring the 
dipolar directional dependence), where $r$ is distance from the broken bond. The probability that a bond at distance $r$
breaks in response to this change in stress is approximately given by $F'(\sigma)c(r,\sigma)$. Thus, the expected 
number of bonds that break in response to stress change due to one bond breaking 
is given by $\lambda \sim \int_1^L rdr F'(\sigma)c(r,\sigma) \sim F'(\sigma)\sigma\log L$.
Substituting the form $F(x) = x^\beta$
gives $\lambda(\sigma,\beta,L) \sim \beta \sigma^\beta \log L = \beta \phi \log L$. 
This shows that $\lim_{\beta \to 0}\lambda(\beta,\phi, L) = 0$ (for fixed $L$), thus there are no avalanches for 
small $\beta$, and the damage is percolation-like. 
For suitable $\beta$ the avalanche progresses as a branching process, where breaking of one bond triggers a few more and so on
($\lambda$ is also known as the branching ratio).
It is well known that integrated avalanche size distribution for such processes is 
a power-law with exponent $\tau_a = 3/2 + 1 = 5/2$; for suitably large $L$
we expect the avalanche size distribution to be a power-law with exponents consistent with the 
mean field value of $5/2$~\cite{zapperi1997}. Finally,
for very large $L$ (or $\beta$), the system flows away from the critical point and the avalanches get cutoff due to 
nucleation effects.
\par
All the ideas discussed so far can be encapsulated neatly in the form of crossover scaling functions.
The scaling form for the cluster size distribution
can be derived by using ideas of scale invariance. Let $G(z_1,\ldots,z_n)$ be a scale invariant 
function, then by definition, $G(\cdot)$ should remain invariant under 
a rescaling by a factor $b$, i.e.~$G(z_1,\ldots,z_n) = b^{\alpha_0}G(z_1b^{\alpha_1},\ldots,z_nb^{\alpha_n})$
for some constants $\alpha_i$. Taking $b = 1 + \epsilon$ and solving up to first order in $\epsilon$
gives the general form of a scale invariant function as 
$G(z_1,\ldots,z_n) = z_1^{-\alpha_0/\alpha_1}\mathcal{G}(z_2 z_1^{-\alpha_2/\alpha_1},\ldots,z_n z_1^{-\alpha_n/\alpha_1})$,
where the universal scaling function, $\mathcal{G}(\cdot)$, and the critical exponents, $\alpha_i/\alpha_1$,
are characteristic of the critical point~\footnote{Other orderings of variables
are equally valid, such as 
$G(z_1,\ldots,z_n) = z_2^{-\alpha_0/\alpha_2}\mathcal{G}_2(z_1 z_2^{-\alpha_1/\alpha_2},\ldots,z_n z_2^{-\alpha_n/\alpha_2})$, etc.
See~\cite{sethna2001} for details.},~\cite{sethna2001}. The variables $z_i$ represent directions
in parameter space near the critical point. The directions with $\alpha_i > 0$ belong to the \emph{relevant parameters}
and those with $\alpha_i < 0$ to \emph{irrelevant parameters}.
We treat $\beta, 1/L$ to be a relevant parameters, and let $u$ be the leading irrelevant parameter 
(the largest of the negative $\alpha_i$).
Thus, ignoring all irrelevant variables but the leading one, the scale invariant distribution of cluster sizes can be written as 
\begin{equation*}
P_c(s|\beta,L) = s^{-\tau_c}\mathcal{F}_c\left(\beta L^{1/\nu_f},sL^{-1/\sigma_c\nu_f},uL^{-\Delta_f/\nu_f}\right)
\end{equation*}
where the subscript $c$ denotes variables associated with the clusters.
We use the subscript $f$ (for fracture) to distinguish the critical exponents from their counterparts in percolation theory.
We know that in the limit of $\beta \to 0$ (at fixed $L$) the cluster size distribution should reduce to 
distribution of percolation clusters at the critical point, thus we can deduce three critical exponent combination,
namely $\tau_c = 187/91 = 2.0549$, $\sigma_c\nu_f = 48/91 = 0.5275$ and $\Delta_f/\nu_f = 72/48 = 1.5$~\cite{Ziff2011}. 
Even thought the clusters created in fracture are \emph{loopless}~\cite{Hansen2012},
the static properties of loopless percolation are identical to usual percolation, thus
the use of percolation critical exponents is valid~\cite{Frank1989}.
The moments of the 
cluster size distribution should scale as (taking a Taylor expansion in $uL^{-\Delta_f/\nu_f}$ for large $L$)
\begin{equation*}
\langle s_c^n \rangle = L^{(n+1 - \tau_c)/\sigma_c\nu_f} \left( \mathcal{J}_n^c(\beta L^{1/\nu_f}) + L^{-\Delta_f/\nu_f} \mathcal{K}_n^c(\beta L^{1/\nu_f}) \right),
\end{equation*}
where $\mathcal{J}^c_n(\cdot),\ \mathcal{K}^c_n(\cdot),\ n = 2, 3\ldots,$ are universal scaling functions~\footnote{This scaling
relation is valid only if $n+1-\tau_c > 0$; we find $\tau_c = 187/91$, thus, $n \geq 2$.}. From a data fitting 
perspective, it is easier to deal with the 
moments (as opposed to the distribution function) because $\mathcal{J}_n^c(\cdot),\ \mathcal{K}_n^c(\cdot)$ are functions
of just one scaling variable. The functions for the avalanche size distribution are completely analogous,
\begin{gather*}
P_a(s|\beta,L) = s^{-\tau_a}\mathcal{F}_a\left(\beta L^{1/\nu_f},sL^{-1/\sigma_a\nu_f},uL^{-\Delta_f/\nu_f}\right),\\
\langle s_a^n \rangle = L^{(n+1 - \tau_a)/\sigma_a\nu_f} \left( \mathcal{J}_n^a(\beta L^{1/\nu_f}) + L^{-\Delta_f/\nu_f} \mathcal{K}_n^a(\beta L^{1/\nu_f}) \right),
\end{gather*}
where $\tau_a$ is expected to be close to its mean field value of $5/2$. 
\begin{figure}[tbp]
\begin{center}
\subfigure[Avalanche size distribution]{\includegraphics[width=0.225\textwidth,angle=0]{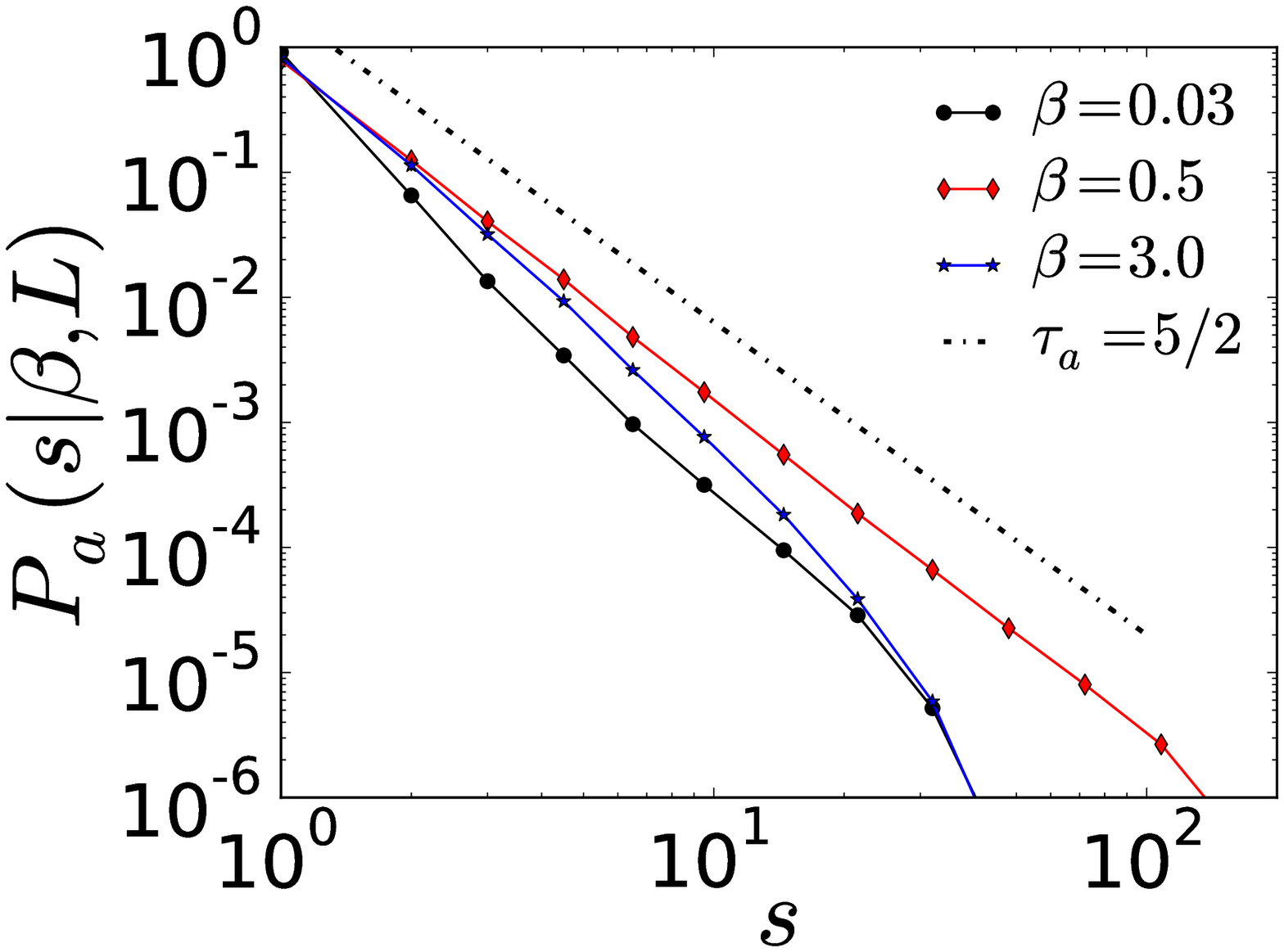}}
\subfigure[Cluster size distribution]{\includegraphics[width=0.225\textwidth,angle=0]{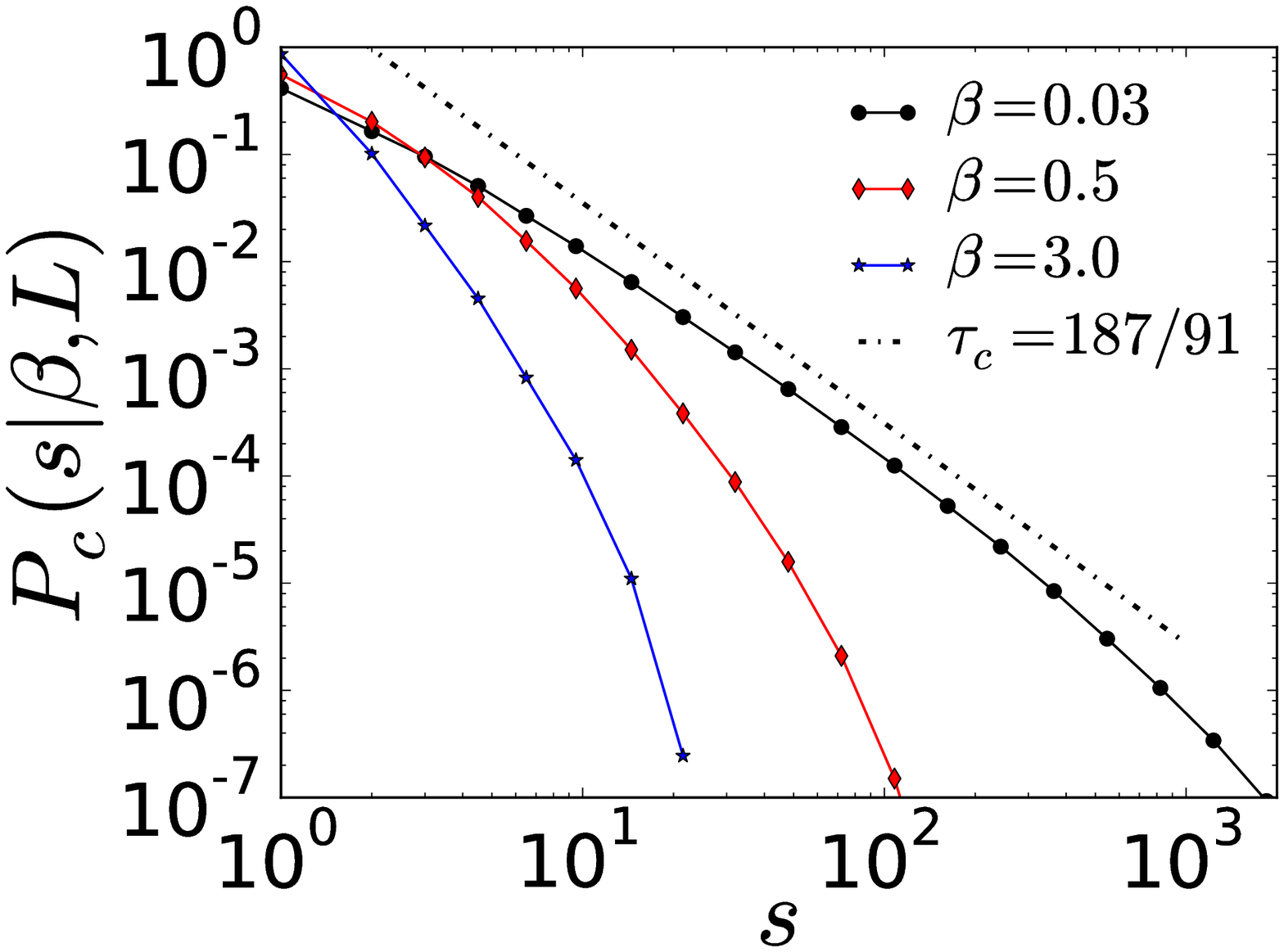}}
\subfigure[Scaling collapse of $\langle s_a^2 \rangle$ ]{\includegraphics[width=0.225\textwidth,angle=0]{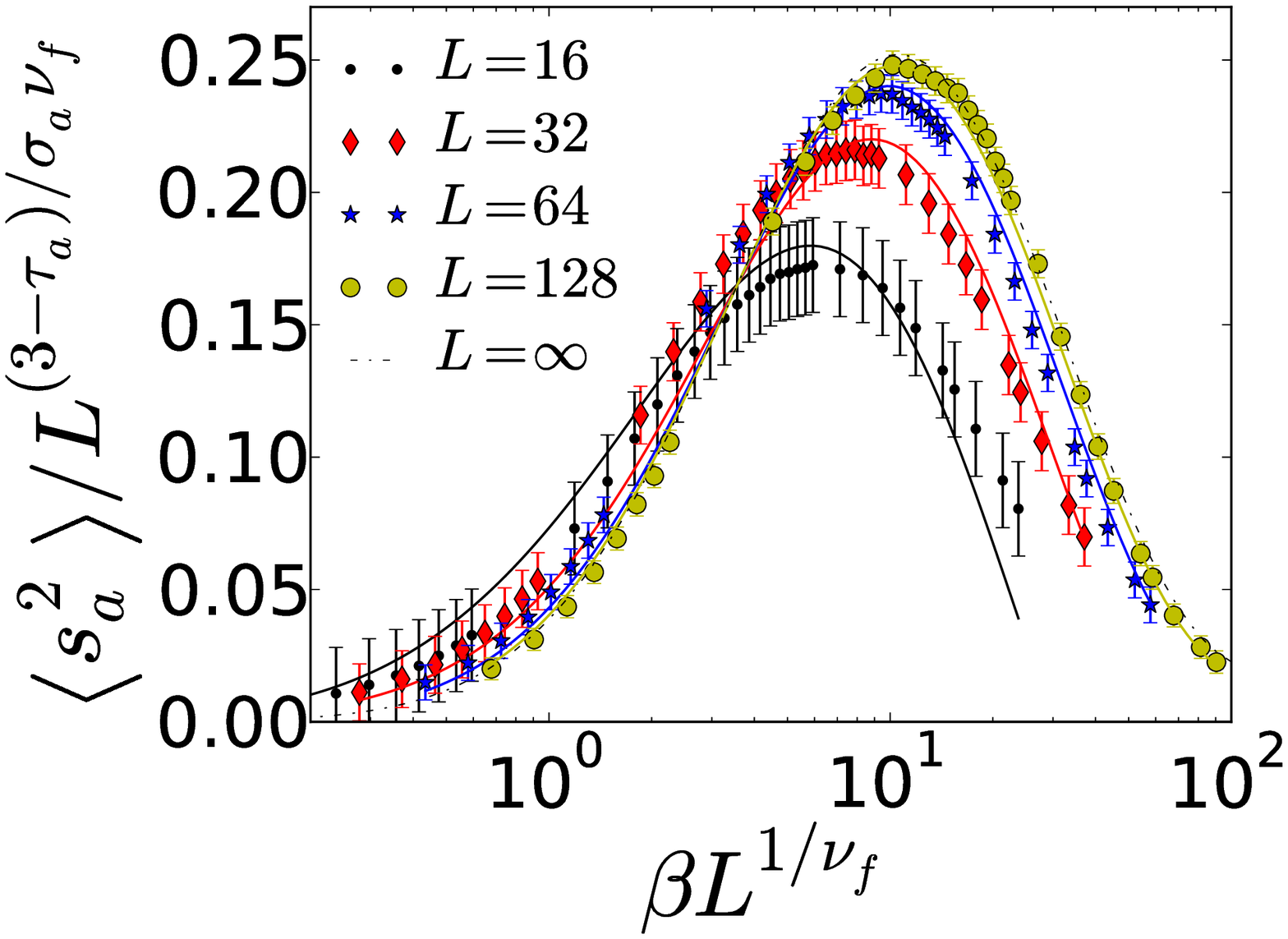}}
\subfigure[Scaling collapse of $\langle s_c^2 \rangle$ ]{\includegraphics[width=0.225\textwidth,angle=0]{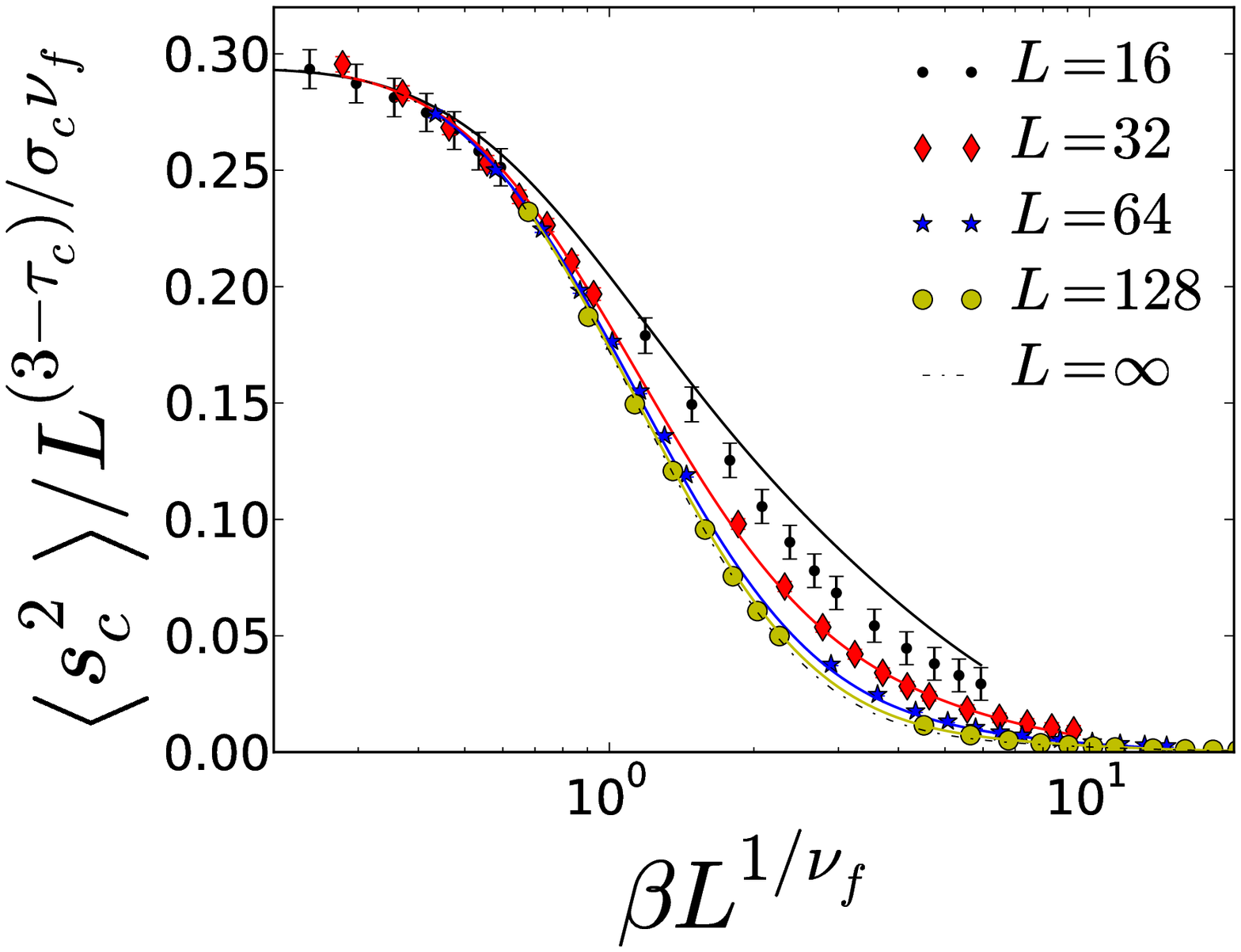}}
\end{center}
\caption{{\bf Scaling theory of fracture. a).} The avalanche size distribution shows a power-law
consistent with the mean field exponent of 5/2 for moderate $\beta$ (= 0.5 at $L=128$). As expected, the power-law
is distorted for much smaller or larger $\beta$. {\bf b).} The cluster size distribution shows a power-law 
that is consistent with the exponent predicted by percolation theory (= 187/91). The power-law cutoff becomes
smaller as one moves away from the critical point.
{\bf c, d).} The scaling forms fit the data well, confirming the predictions of the scaling theory.
Higher moments of the distributions fit the scaling forms as well (not shown here). Notice the 
significant finite-size effects as the data gets closer to the $L = \infty $ curve with 
increasing system sizes.
} 
\label{fig:Fits}
\end{figure}
\par
We have done numerical simulations to verify our theoretical predictions. 
We did extensive statistical sampling of systems of size up to $L=128$ and $\beta$ between 0.03 and 8.
In order to fit the data to the scaling predictions we use the following functional forms for the scaling
functions for the moments of the cluster size distribution (with $y_c(x) \equiv (\log x - \mu_c)/\alpha_c$)
\begin{align*}
\mathcal{J}_n^c(x) & = a_{0,n} \mathrm{erf}\left( y_c(x)\right) + e^{ -\left(y_c(x)\right)^2} \sum_{i=0}^{i=m}A^c_{i,n}H_i\left( y_c(x) \right),\\
\mathcal{K}_n^c(x) &= a_{1,n}\mathrm{erf}\left( y_c(x)\right) + e^{ -\left(y_c(x)\right)^2} \sum_{i=0}^{i=m}B^c_{i,n}H_i\left( y_c(x) \right),
\end{align*}
where 
$\mu_c,\ \alpha_c,\ a_{0,n},\ a_{1,n},\ A_{i,n}^c,\ B_{i,n}^c$ are fitting parameters, 
$\mathrm{erf}(\cdot)$ is the error function, 
and $H_i(\cdot)$ is the $i^{th}$ Hermite polynomial. We use the first three Hermite polynomials 
in the expansion, i.e., $m = 3$.
The corresponding forms for the avalanches are (with $y_a(x) \equiv (\log x - \mu_a)/\alpha_a$)
\begin{align*}
\mathcal{J}_n^a(x) &= e^{ -\left(y_a(x)\right)^2} \sum_{i=0}^{i=m}A^a_{i,n}H_i\left( y_a(x)\right),\\
\mathcal{K}_n^a(x) &= e^{ -\left(y_a(x)\right)^2} \sum_{i=0}^{i=m}B^a_{i,n}H_i\left( y_a(x) \right).
\end{align*}

\begin{figure}[tbp]
\begin{center}
\includegraphics[width=0.4\textwidth,angle=0]{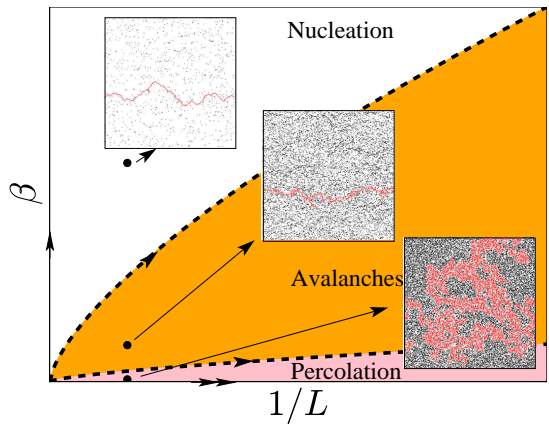}
\end{center}
\caption{{\bf Phase diagram for brittle fracture in disordered media~\cite{hansen2006}.} 
Disorder decreases along the $\beta$ axis; nucleation governs the behavior
for small disorder or long length scales. Percolation
is characteristic of the large disorder regime, while the crossover
region exhibits interesting critical behavior in the form of
scale free distributions of avalanche sizes. The topology 
of the fractured samples evolves from percolation-like damage
for large disorder to well defined sharp cracks in the nucleated regime.
The phase boundaries are quantitatively somewhat arbitrary, and are set 
at the value of the scaling variable $\beta L^{1/\nu_f}$ at which 
the second moment of the avalanche size become half of its peak value (for the avalanche 
phase); the boundary of the percolation phase is found analogously.
} 
\label{fig:PhaseDiagram}
\end{figure}
\par
The forms of the scaling functions are chosen so that they have the correct asymptotic behavior. As discussed previously,
we know that $\lim_{\beta L^{1/\nu_f} \to 0, \infty} \mathcal{J}_n^a(\beta L^{1/\nu_f}) = 0$ since there are no 
avalanches for very small $\beta$ (at fixed $L$) and at very large $L$ (at fixed $\beta$). On the other hand
we know $\lim_{\beta L^{1/\nu_f} \to 0} \mathcal{J}_n^c(\beta L^{1/\nu_f}) = C$, for some constant $C$ (according to 
percolation theory) and $\lim_{\beta L^{1/\nu_f} \to \infty} \mathcal{J}_n^c(\beta L^{1/\nu_f}) = 0$ since 
there are no clusters in the nucleation dominated regime away from the critical point. The forms used here
satisfy all these requirements.
\par
Figure~\ref{fig:Fits} shows the size distributions as well as fits to the scaling forms. It is evident that 
the data is consistent with the scaling theory. Based on joint fits for $n = 2,\ 3$ 
($n = 3$ not shown in figure~\ref{fig:Fits}) we estimate the following 
values of the critical exponents: 
$\nu_f = 1.56 \pm 0.30$, $\sigma_a = 0.47 \pm 0.15$,
$\Delta_f = 2.35 \pm 1.50$, $\sigma_c = 0.34 \pm 0.08$. 
For the fits shown in figure~\ref{fig:Fits} the exponent $\tau_a$ is held 
at its mean-field value of 5/2, while unbiased fits yield $\tau_a = 2.45 \pm 0.25$.
The scaling exponent
$\tau_c$, and the exponent combinations $\sigma_c\nu_f$, $\Delta_f/\nu_f$ are held at their theoretical values of $187/91$
and 48/91, 72/48, respectively.
The statistical error bars are much smaller than the error bars reported here.
We have estimated the error bars due to systematic errors by using a variety of techniques such as varying 
the number of terms in the scaling functions, trying different fitting forms, varying 
the critical range for the fits, varying the error bars on the data over a reasonable range, etc.
Figure~\ref{fig:PhaseDiagram} shows the phase diagram that emerges from our analysis~\footnote{%
Ref.~\cite{hansen2006} has foreshadowed our `phase diagram' of figure~\ref{fig:PhaseDiagram}, with 
their `diffuse phase' (see figure 4 of Ref.~\cite{hansen2006}) 
corresponding to our avalanche regime, but their results were not based a scaling description. Indeed, our 
crossover analysis is inconsistent with their phase boundary (or any phase boundary as $1/L \to 0$)}.
In the $\beta - 1/L$ space,
curves along which the scaling variable $\beta L^{1/\nu_f}$ attains a critical value demarcate the boundary between
qualitatively different behavior. Note that the exact position of the boundaries is somewhat arbitrary, since this 
is a not an abrupt (first order) transition; however, the diagram is qualitatively accurate.

\par
The critical phenomena associated with fracture has several intriguing characteristics. Firstly, the 
scaling function associated with the avalanches has a singularity at 0, 
$\lim_{\beta L^{1/\nu_f}\to 0} \mathcal{J}_n^a(\beta L^{1/\nu_f}) = 0$, that subdues the avalanche behavior 
as the critical point is approached. Secondly, there is no point in the phase diagram (except for 
the $\beta = 0$ limit) that shows any critical phenomena in the 
limit of $L\to \infty$. Thus, scale invariance itself becomes a finite-size effect; perhaps 
this phenomena should be named \emph{finite-sized criticality}. Finally,
it is rather remarkable that the critical phenomena (typically associated
with continuous phase transitions) gives way to nucleation (a first order
transition) in the limit of long length scales! Thus, fracture 
has mixed first order and continuous transition character.
Transitions of mixed first order and continuous character have become somewhat of a theme
in the past decade or so. Recently we noted that the Mott transition and dielectric 
breakdown have a mixed character~\cite{shekhawat2011}; similar findings have been reported 
in a variety of fields such as jamming transitions, rigidity percolation~\cite{mao2011}, and phase-separated manganites
~\cite{parisi2004}.
\par
In conclusion, we have presented a scaling theory of fracture that builds on 
renormalization group ideas and unifies several disparate results in the 
field. Our theory shows that percolation-like behavior as well as 
the scale invariant precursor avalanches leading to fracture are finite-size effects. We show
that on long length scales brittle fracture is always nucleated. We hope
that our analysis will pave the way for a deeper understanding of 
the many mysteries associated with the phenomenon of fracture.

\begin{acknowledgments}
We would like to thank S.~L.~Phoenix for insightful discussions. A.S.~and J.P.S.~were supported by DOE-BES
DE-FG02-07ER46393. S.Z.~acknowledges financial support from ERC-AdG-2011 SIZEFFECT. 
This research was supported in part by the 
National Science Foundation through TeraGrid resources provided by the 
Louisiana Optical Network Initiative (LONI) under grant number TG-DMR100025.
\end{acknowledgments}
\bibliography{fractureI}

\begin{thebibliography}{29}
\expandafter\ifx\csname natexlab\endcsname\relax\def\natexlab#1{#1}\fi
\expandafter\ifx\csname bibnamefont\endcsname\relax
  \def\bibnamefont#1{#1}\fi
\expandafter\ifx\csname bibfnamefont\endcsname\relax
  \def\bibfnamefont#1{#1}\fi
\expandafter\ifx\csname citenamefont\endcsname\relax
  \def\citenamefont#1{#1}\fi
\expandafter\ifx\csname url\endcsname\relax
  \def\url#1{\texttt{#1}}\fi
\expandafter\ifx\csname urlprefix\endcsname\relax\def\urlprefix{URL }\fi
\providecommand{\bibinfo}[2]{#2}
\providecommand{\eprint}[2][]{\url{#2}}

\bibitem[{\citenamefont{Roux et~al.}(1988)\citenamefont{Roux, Hansen, Herrmann,
  and Guyon}}]{roux1988}
\bibinfo{author}{\bibfnamefont{S.}~\bibnamefont{Roux}},
  \bibinfo{author}{\bibfnamefont{A.}~\bibnamefont{Hansen}},
  \bibinfo{author}{\bibfnamefont{H.}~\bibnamefont{Herrmann}}, \bibnamefont{and}
  \bibinfo{author}{\bibfnamefont{E.}~\bibnamefont{Guyon}},
  \bibinfo{journal}{Journal of Statistical Physics}
  \textbf{\bibinfo{volume}{52}}, \bibinfo{pages}{237} (\bibinfo{year}{1988}).

\bibitem[{\citenamefont{Hansen and Schmittbuhl}(2003)}]{hansen2003}
\bibinfo{author}{\bibfnamefont{A.}~\bibnamefont{Hansen}} \bibnamefont{and}
  \bibinfo{author}{\bibfnamefont{J.}~\bibnamefont{Schmittbuhl}},
  \bibinfo{journal}{Phys. Rev. Lett.} \textbf{\bibinfo{volume}{90}}
  (\bibinfo{year}{2003}).

\bibitem[{\citenamefont{Hansen et~al.}(1991)\citenamefont{Hansen, Hinrichsen,
  and Roux}}]{hansen1991}
\bibinfo{author}{\bibfnamefont{A.}~\bibnamefont{Hansen}},
  \bibinfo{author}{\bibfnamefont{E.~L.} \bibnamefont{Hinrichsen}},
  \bibnamefont{and} \bibinfo{author}{\bibfnamefont{S.}~\bibnamefont{Roux}},
  \bibinfo{journal}{Phys. Rev. B} \textbf{\bibinfo{volume}{43}},
  \bibinfo{pages}{665} (\bibinfo{year}{1991}).

\bibitem[{\citenamefont{Toussaint and Hansen}(2006)}]{hansen2006}
\bibinfo{author}{\bibfnamefont{R.}~\bibnamefont{Toussaint}} \bibnamefont{and}
  \bibinfo{author}{\bibfnamefont{A.}~\bibnamefont{Hansen}},
  \bibinfo{journal}{Phys. Rev. E} \textbf{\bibinfo{volume}{73}},
  \bibinfo{pages}{046103} (\bibinfo{year}{2006}).

\bibitem[{\citenamefont{Zapperi et~al.}(1997)\citenamefont{Zapperi, Ray,
  Stanley, and Vespignani}}]{zapperi1997}
\bibinfo{author}{\bibfnamefont{S.}~\bibnamefont{Zapperi}},
  \bibinfo{author}{\bibfnamefont{P.}~\bibnamefont{Ray}},
  \bibinfo{author}{\bibfnamefont{H.~E.} \bibnamefont{Stanley}},
  \bibnamefont{and}
  \bibinfo{author}{\bibfnamefont{A.}~\bibnamefont{Vespignani}},
  \bibinfo{journal}{Phys. Rev. Lett.} \textbf{\bibinfo{volume}{78}}
  (\bibinfo{year}{1997}).

\bibitem[{\citenamefont{Duxbury et~al.}(1987)\citenamefont{Duxbury, Beale, and
  Leath}}]{duxbury1987}
\bibinfo{author}{\bibfnamefont{P.~M.} \bibnamefont{Duxbury}},
  \bibinfo{author}{\bibfnamefont{P.~D.} \bibnamefont{Beale}}, \bibnamefont{and}
  \bibinfo{author}{\bibfnamefont{P.~L.} \bibnamefont{Leath}},
  \bibinfo{journal}{Phys. Rev. B} \textbf{\bibinfo{volume}{36}}
  (\bibinfo{year}{1987}).

\bibitem[{\citenamefont{Duxbury et~al.}(1986)\citenamefont{Duxbury, Beale, and
  Leath}}]{duxbury1986}
\bibinfo{author}{\bibfnamefont{P.~M.} \bibnamefont{Duxbury}},
  \bibinfo{author}{\bibfnamefont{P.~D.} \bibnamefont{Beale}}, \bibnamefont{and}
  \bibinfo{author}{\bibfnamefont{P.~L.} \bibnamefont{Leath}},
  \bibinfo{journal}{Phys. Rev. Lett.} \textbf{\bibinfo{volume}{57}}
  (\bibinfo{year}{1986}).

\bibitem[{\citenamefont{Manzato et~al.}(2012)\citenamefont{Manzato, Shekhawat,
  Nukala, Alava, Sethna, and Zapperi}}]{shekhawat2012}
\bibinfo{author}{\bibfnamefont{C.}~\bibnamefont{Manzato}},
  \bibinfo{author}{\bibfnamefont{A.}~\bibnamefont{Shekhawat}},
  \bibinfo{author}{\bibfnamefont{P.~K. V.~V.} \bibnamefont{Nukala}},
  \bibinfo{author}{\bibfnamefont{M.~J.} \bibnamefont{Alava}},
  \bibinfo{author}{\bibfnamefont{J.~P.} \bibnamefont{Sethna}},
  \bibnamefont{and} \bibinfo{author}{\bibfnamefont{S.}~\bibnamefont{Zapperi}},
  \bibinfo{journal}{Phys. Rev. Lett.} \textbf{\bibinfo{volume}{108}},
  \bibinfo{pages}{065504} (\bibinfo{year}{2012}).

\bibitem[{\citenamefont{Chakrabarti and Benguigui}(1997)}]{chakrabarti}
\bibinfo{author}{\bibfnamefont{B.~K.} \bibnamefont{Chakrabarti}}
  \bibnamefont{and} \bibinfo{author}{\bibfnamefont{L.~G.}
  \bibnamefont{Benguigui}}, \emph{\bibinfo{title}{Statistical Physics of
  Fracture and Breakdown in Disordered Systems}} (\bibinfo{publisher}{Oxford
  Science Publications, Oxford}, \bibinfo{year}{1997}).

\bibitem[{\citenamefont{Bonamy and Bouchaud}(2011)}]{bonamy11}
\bibinfo{author}{\bibfnamefont{D.}~\bibnamefont{Bonamy}} \bibnamefont{and}
  \bibinfo{author}{\bibfnamefont{E.}~\bibnamefont{Bouchaud}},
  \bibinfo{journal}{Physics Reports} \textbf{\bibinfo{volume}{498}},
  \bibinfo{pages}{1 } (\bibinfo{year}{2011}).

\bibitem[{\citenamefont{Hemmer and Hansen}(1992)}]{hemmer1992}
\bibinfo{author}{\bibfnamefont{P.~C.} \bibnamefont{Hemmer}} \bibnamefont{and}
  \bibinfo{author}{\bibfnamefont{A.}~\bibnamefont{Hansen}},
  \bibinfo{journal}{Journal of applied mechanics}
  \textbf{\bibinfo{volume}{59}}, \bibinfo{pages}{909} (\bibinfo{year}{1992}).

\bibitem[{\citenamefont{Petri et~al.}(1994)\citenamefont{Petri, Paparo,
  Vespignani, Alippi, and Costantini}}]{petri94}
\bibinfo{author}{\bibfnamefont{A.}~\bibnamefont{Petri}},
  \bibinfo{author}{\bibfnamefont{G.}~\bibnamefont{Paparo}},
  \bibinfo{author}{\bibfnamefont{A.}~\bibnamefont{Vespignani}},
  \bibinfo{author}{\bibfnamefont{A.}~\bibnamefont{Alippi}}, \bibnamefont{and}
  \bibinfo{author}{\bibfnamefont{M.}~\bibnamefont{Costantini}},
  \bibinfo{journal}{Phys. Rev. Lett.} \textbf{\bibinfo{volume}{73}},
  \bibinfo{pages}{3423} (\bibinfo{year}{1994}).

\bibitem[{\citenamefont{Garcimartin et~al.}(1997)\citenamefont{Garcimartin,
  Guarino, Bellon, and Ciliberto}}]{garcimartin97}
\bibinfo{author}{\bibfnamefont{A.}~\bibnamefont{Garcimartin}},
  \bibinfo{author}{\bibfnamefont{A.}~\bibnamefont{Guarino}},
  \bibinfo{author}{\bibfnamefont{L.}~\bibnamefont{Bellon}}, \bibnamefont{and}
  \bibinfo{author}{\bibfnamefont{S.}~\bibnamefont{Ciliberto}},
  \bibinfo{journal}{Phys. Rev. Lett.} \textbf{\bibinfo{volume}{79}},
  \bibinfo{pages}{3202} (\bibinfo{year}{1997}).

\bibitem[{\citenamefont{Salminen et~al.}(2002)\citenamefont{Salminen, Tolvanen,
  and Alava}}]{salminen02}
\bibinfo{author}{\bibfnamefont{L.~I.} \bibnamefont{Salminen}},
  \bibinfo{author}{\bibfnamefont{A.~I.} \bibnamefont{Tolvanen}},
  \bibnamefont{and} \bibinfo{author}{\bibfnamefont{M.~J.} \bibnamefont{Alava}},
  \bibinfo{journal}{Phys. Rev. Lett.} \textbf{\bibinfo{volume}{89}},
  \bibinfo{pages}{185503} (\bibinfo{year}{2002}).

\bibitem[{\citenamefont{Alava et~al.}(2006)\citenamefont{Alava, Nukala, and
  Zapperi}}]{alava2006}
\bibinfo{author}{\bibfnamefont{M.~J.} \bibnamefont{Alava}},
  \bibinfo{author}{\bibfnamefont{P.~K. V.~V.} \bibnamefont{Nukala}},
  \bibnamefont{and} \bibinfo{author}{\bibfnamefont{S.}~\bibnamefont{Zapperi}},
  \bibinfo{journal}{Advances in Physics} \textbf{\bibinfo{volume}{55}}
  (\bibinfo{year}{2006}).

\bibitem[{\citenamefont{Bazant}(2004)}]{bazant04b}
\bibinfo{author}{\bibfnamefont{Z.~P.} \bibnamefont{Bazant}},
  \bibinfo{journal}{PNAS} \textbf{\bibinfo{volume}{101}},
  \bibinfo{pages}{13400} (\bibinfo{year}{2004}).

\bibitem[{\citenamefont{Harlow and Phoenix}(1978)}]{harlow1978}
\bibinfo{author}{\bibfnamefont{D.~G.} \bibnamefont{Harlow}} \bibnamefont{and}
  \bibinfo{author}{\bibfnamefont{S.~L.} \bibnamefont{Phoenix}},
  \bibinfo{journal}{Journal of composite materials}
  \textbf{\bibinfo{volume}{12}}, \bibinfo{pages}{195} (\bibinfo{year}{1978}).

\bibitem[{\citenamefont{Hansen and Hemmer}(1994)}]{hansen94}
\bibinfo{author}{\bibfnamefont{A.}~\bibnamefont{Hansen}} \bibnamefont{and}
  \bibinfo{author}{\bibfnamefont{P.~C.} \bibnamefont{Hemmer}},
  \bibinfo{journal}{Phys. Lett. A} \textbf{\bibinfo{volume}{184}},
  \bibinfo{pages}{394} (\bibinfo{year}{1994}).

\bibitem[{\citenamefont{Zapperi
  et~al.}(2005{\natexlab{a}})\citenamefont{Zapperi, Nukala, and
  Simunovic}}]{zapperi05}
\bibinfo{author}{\bibfnamefont{S.}~\bibnamefont{Zapperi}},
  \bibinfo{author}{\bibfnamefont{P.~K. V.~V.} \bibnamefont{Nukala}},
  \bibnamefont{and}
  \bibinfo{author}{\bibfnamefont{S.}~\bibnamefont{Simunovic}},
  \bibinfo{journal}{Phys. Rev. E} \textbf{\bibinfo{volume}{71}},
  \bibinfo{pages}{026106} (\bibinfo{year}{2005}{\natexlab{a}}).

\bibitem[{\citenamefont{Zapperi
  et~al.}(2005{\natexlab{b}})\citenamefont{Zapperi, Nukala, and
  Simunovic}}]{zapperi05b}
\bibinfo{author}{\bibfnamefont{S.}~\bibnamefont{Zapperi}},
  \bibinfo{author}{\bibfnamefont{P.~K. V.~V.} \bibnamefont{Nukala}},
  \bibnamefont{and}
  \bibinfo{author}{\bibfnamefont{S.}~\bibnamefont{Simunovic}},
  \bibinfo{journal}{Physica A} \textbf{\bibinfo{volume}{357}},
  \bibinfo{pages}{129} (\bibinfo{year}{2005}{\natexlab{b}}).

\bibitem[{\citenamefont{Nukala and Simunovic}(2003)}]{nukala2003}
\bibinfo{author}{\bibfnamefont{P.~K. V.~V.} \bibnamefont{Nukala}}
  \bibnamefont{and}
  \bibinfo{author}{\bibfnamefont{S.}~\bibnamefont{Simunovic}},
  \bibinfo{journal}{J. Phys. A: Math. Gen.} \textbf{\bibinfo{volume}{36}},
  \bibinfo{pages}{11403} (\bibinfo{year}{2003}).

\bibitem[{\citenamefont{Kahng et~al.}(1988)\citenamefont{Kahng, Batrouni,
  Redner, de~Arcangelis, and Herrmann}}]{kahng88}
\bibinfo{author}{\bibfnamefont{B.}~\bibnamefont{Kahng}},
  \bibinfo{author}{\bibfnamefont{G.~G.} \bibnamefont{Batrouni}},
  \bibinfo{author}{\bibfnamefont{S.}~\bibnamefont{Redner}},
  \bibinfo{author}{\bibfnamefont{L.}~\bibnamefont{de~Arcangelis}},
  \bibnamefont{and} \bibinfo{author}{\bibfnamefont{H.~J.}
  \bibnamefont{Herrmann}}, \bibinfo{journal}{Phys. Rev. B}
  \textbf{\bibinfo{volume}{37(13)}}, \bibinfo{pages}{7625}
  (\bibinfo{year}{1988}).

\bibitem[{\citenamefont{Moreira et~al.}(2012)\citenamefont{Moreira, Oliveira,
  Hansen, Araujo, Herrmann, and J.~S.~Andrade}}]{Hansen2012}
\bibinfo{author}{\bibfnamefont{A.~A.} \bibnamefont{Moreira}},
  \bibinfo{author}{\bibfnamefont{C.~L.~N.} \bibnamefont{Oliveira}},
  \bibinfo{author}{\bibfnamefont{A.}~\bibnamefont{Hansen}},
  \bibinfo{author}{\bibfnamefont{N.~A.~M.} \bibnamefont{Araujo}},
  \bibinfo{author}{\bibfnamefont{H.~J.} \bibnamefont{Herrmann}},
  \bibnamefont{and}
  \bibinfo{author}{\bibfnamefont{J.}~\bibnamefont{J.~S.~Andrade}},
  \bibinfo{journal}{Arxiv preprint arXiv:1206.1233}  (\bibinfo{year}{2012}).

\bibitem[{\citenamefont{Sethna et~al.}(2001)\citenamefont{Sethna, Dahmen, and
  Myers}}]{sethna2001}
\bibinfo{author}{\bibfnamefont{J.~P.} \bibnamefont{Sethna}},
  \bibinfo{author}{\bibfnamefont{K.~A.} \bibnamefont{Dahmen}},
  \bibnamefont{and} \bibinfo{author}{\bibfnamefont{C.~R.} \bibnamefont{Myers}},
  \bibinfo{journal}{Nature} \textbf{\bibinfo{volume}{410}},
  \bibinfo{pages}{242} (\bibinfo{year}{2001}).

\bibitem[{\citenamefont{Ziff}(2011)}]{Ziff2011}
\bibinfo{author}{\bibfnamefont{R.~M.} \bibnamefont{Ziff}},
  \bibinfo{journal}{Phys. Rev. E} \textbf{\bibinfo{volume}{83}},
  \bibinfo{pages}{020107} (\bibinfo{year}{2011}).

\bibitem[{\citenamefont{Tzschichholz et~al.}(1989)\citenamefont{Tzschichholz,
  Bunde, and Havlin}}]{Frank1989}
\bibinfo{author}{\bibfnamefont{F.}~\bibnamefont{Tzschichholz}},
  \bibinfo{author}{\bibfnamefont{A.}~\bibnamefont{Bunde}}, \bibnamefont{and}
  \bibinfo{author}{\bibfnamefont{S.}~\bibnamefont{Havlin}},
  \bibinfo{journal}{Phys. Rev. A} \textbf{\bibinfo{volume}{39}},
  \bibinfo{pages}{5470} (\bibinfo{year}{1989}).

\bibitem[{\citenamefont{Shekhawat et~al.}(2011)\citenamefont{Shekhawat,
  Papanikolaou, Zapperi, and Sethna}}]{shekhawat2011}
\bibinfo{author}{\bibfnamefont{A.}~\bibnamefont{Shekhawat}},
  \bibinfo{author}{\bibfnamefont{S.}~\bibnamefont{Papanikolaou}},
  \bibinfo{author}{\bibfnamefont{S.}~\bibnamefont{Zapperi}}, \bibnamefont{and}
  \bibinfo{author}{\bibfnamefont{J.~P.} \bibnamefont{Sethna}},
  \bibinfo{journal}{Physical Review Letters} \textbf{\bibinfo{volume}{107}},
  \bibinfo{pages}{276401} (\bibinfo{year}{2011}).

\bibitem[{\citenamefont{Ellenbroek and Mao}(2011)}]{mao2011}
\bibinfo{author}{\bibfnamefont{W.~G.} \bibnamefont{Ellenbroek}}
  \bibnamefont{and} \bibinfo{author}{\bibfnamefont{X.}~\bibnamefont{Mao}},
  \bibinfo{journal}{EPL (Europhysics Letters)} \textbf{\bibinfo{volume}{96}},
  \bibinfo{pages}{54002} (\bibinfo{year}{2011}).

\bibitem[{\citenamefont{Ghivelder et~al.}(2004)\citenamefont{Ghivelder,
  Freitas, das Virgens, Continentino, Martinho, Granja, Quintero, Leyva., Levy,
  and Parisi}}]{parisi2004}
\bibinfo{author}{\bibfnamefont{L.}~\bibnamefont{Ghivelder}},
  \bibinfo{author}{\bibfnamefont{R.~S.} \bibnamefont{Freitas}},
  \bibinfo{author}{\bibfnamefont{M.~G.} \bibnamefont{das Virgens}},
  \bibinfo{author}{\bibfnamefont{M.~A.} \bibnamefont{Continentino}},
  \bibinfo{author}{\bibfnamefont{H.}~\bibnamefont{Martinho}},
  \bibinfo{author}{\bibfnamefont{L.}~\bibnamefont{Granja}},
  \bibinfo{author}{\bibfnamefont{M.}~\bibnamefont{Quintero}},
  \bibinfo{author}{\bibfnamefont{G.}~\bibnamefont{Leyva.}},
  \bibinfo{author}{\bibfnamefont{P.}~\bibnamefont{Levy}}, \bibnamefont{and}
  \bibinfo{author}{\bibfnamefont{F.}~\bibnamefont{Parisi}},
  \bibinfo{journal}{Phys. Rev. B} \textbf{\bibinfo{volume}{69}},
  \bibinfo{pages}{214414} (\bibinfo{year}{2004}).

\end{thebibliography}
\end{document}